\documentclass[12pt]{article}
\newcommand{\bce}{\begin{center}}
\newcommand{\ece}{\end{center}}
\newcommand{\beq}{\begin{equation}}
\newcommand{\eeq}{\end{equation}}
\newcommand{\bea}{\vspace{0.25cm}\begin{eqnarray}}
\newcommand{\eea}{\end{eqnarray}}

\newcommand{\ba}{\begin{array}}
\newcommand{\ea}{\end{array}}

\newcommand{\doublespace}{
    \renewcommand{\baselinestretch}{1.6}\large\normalsize}

\def\lsim{\mathrel{\rlap{\lower4pt\hbox{\hskip1pt$\sim$}}
    \raise1pt\hbox{$<$}}}     
\def\gsim{\mathrel{\rlap{\lower4pt\hbox{\hskip1pt$\sim$}}
    \raise1pt\hbox{$>$}}}     

\def\lsim{\mathrel{\rlap{\lower4pt\hbox{\hskip1pt$\sim$}}
    \raise1pt\hbox{$<$}}}         
\def\gsim{\mathrel{\rlap{\lower4pt\hbox{\hskip1pt$\sim$}}
    \raise1pt\hbox{$>$}}}         

\def\beq{\begin{equation}}
\def\endeq{\end{equation}}
\def\arr{\begin{eqnarray}}
\def\endarr{\end{eqnarray}}

\makeindex
\begin{document}
\vspace{2cm}
\begin{center}
{\bf \huge Universal pattern in (e,e'p) at large
missing momenta: quasi-deuteron or diffractive
final state interactions? \\}
\vspace{1cm}
{\bf A.Bianconi$^{1,2)}$, S.Jeschonnek$^{3)}$,
N.N.Nikolaev$^{3,4)}$, B.G.Zakharov$^{3,4}$ } \medskip\\
{\small \sl
$^{1)}$Istituto Nazionale di Fisica Nucleare,
Sezione di Pavia, Pavia, Italy \\
$^{2)}$Dipartimento Fisica Nucleare e Teorica,
Universit\`a di Pavia, Italy \\
$^{3)}$IKP(Theorie), Forschungszentrum  J\"ulich GmbH.,\\
D-52425 J\"ulich, Germany \\
$^{4)}$L.D.Landau Institute for Theoretical Physics, \\
GSP-1, 117940, ul.Kosygina 2, V-334 Moscow, Russia
\vspace{1cm}\\}
{\bf \LARGE A b s t r a c t \bigskip\\}
\end{center}
The intrinsic single particle
momentum distributions in nuclei are supposed
to show a universal behavior at large momenta, dominated
by short-range correlated pairs, or quasi-deuterons.
We discuss whether the quasi-deuteron universality
survives the final state interaction
 effects, which are present in the missing
momentum spectra measured
in $A(e,e'p)$ experiments at GeV energies.
We demonstrate that in the observed missing momentum spectra
an approximate universality is present, but
originating from the universal pattern of
diffractive final state interactions of the struck proton
independent of the target nucleus.

\medskip

\newpage
\doublespace

In the past decades, a great theoretical effort has been devoted to
the study of the nuclear single particle momentum distribution
$N(k)$. Particular importance is attributed to the large momentum tail
of $N(k)$, which is expected to be dominated by two nucleon
short-range correlations, or ``quasi-deuteron'' configurations
\cite{reviews} (QD onwards).  The quasi-deuteron idea was first
discussed quantitatively in \cite{levinger} (see also \cite{zabo}).
This name originates in the predicted analogous behavior of
short-range interactions in all nuclei, from deuteron to nuclear
matter.  This should be reflected in a similar shape for the large-$k$
tail ($k > 1.5\div 2$ fm$^{-1}$) of $N(k)$ in deuteron or in any other
nucleus (hereafter the universality of $N(k)$ refers to the
$k$-dependence, apart from a $k$-independent normalization factor).
An example can be seen, e.g., in Fig.5 of \cite{wiringa92} where the
ratio between $^4$He and $D$ distributions varies roughly between 2.5
and 4.5 in the range 1 fm$^{-1} < k <$ 4 fm$^{-1}$.

However, the pressing issue is how this important intrinsic feature of
the nuclear structure can be tested in the missing momentum
distribution $W(\vec{p}_{m})$ measured experimentally in $(e,e'p)$
reactions at large missing momentum $\vec{p}_{m}$.  In two previous
papers \cite{Helium4,Deuter} we have shown that final state
interactions (FSI) between the struck nucleon and the spectator
nucleons take over at large $\vec{p}_{m}$, making the observed
$W(\vec{p}_{m})$ substantially different from the ground state
distribution $N(k)$.  Furthermore, the sensitivity to the details of
the nuclear structure is lost to a large extent.  In this
communication we wish to focus on an approximate universality of the
observed $W(\vec{p}_{m})$ for $^{4}He$ and $D$ targets, which is
driven by the target independence of FSI between the struck and
spectator nucleons.

We confine ourselves to large $Q^{2}$ and high kinetic energy of the
struck proton $T_{kin}\approx Q^{2}/2m_{p}$.  The very nature of
nucleon-nucleon interaction changes from the purely elastic potential
scattering at low energies to a strongly absorptive, diffractive small
angle scattering at $T_{kin}\gsim $ 0.5-1\,GeV.  In this high energy
regime, the Compton wavelength of the struck proton is much smaller
than the size of a nucleon and/or average nucleon-nucleon separation
in nuclei.  The Glauber model \cite{Glauber} becomes a natural
framework for quantitative description of FSI.

The single particle momentum distribution $N(k)$ coincides
with the longitudinal reduced response in (e,e'p)
scattering in the Plane Wave Impulse Approximation
(i.e. neglecting FSI) integrated over all the missing
energies. This can be written as:
\begin{equation}
N(p_{m})\ =\ \sum_n
\Big\vert\big\langle\Psi_n\big\vert e^{i\vec{p}_{m}\vec r}
\big\vert\Psi_A\big\rangle\Big\vert^2
\end{equation}
where $\Psi_A$ is the ground-state wave function of the target nucleus
(mass number A), the sum goes over all the possible states $\Psi_n$
for the recoiling system of A-1 particles, $\vec r$ is the position of
the struck proton, $\vec{p}_{m}$ $\equiv$ $\vec q - \vec p$ is the
missing momentum (in PWIA, the missing momentum $p_m$ coincides with
the initial nucleon momentum $k$), where $\vec q$ is the momentum
transfer from the electron to the target and $\vec p$ is the momentum
of the detected proton.  The angle between $\vec p_m$ and $\vec q$ is
denoted by $\theta$.

In the calculation of the experimentally measured (e,e'p) coincidence
distribution one must include FSI distortions,
\begin{equation}
W(\vec p_m)\ =\ \sum_n
\Big\vert\big\langle\Psi_n\big\vert e^{i\vec p_m\vec r}S(1,...A)
\big\vert\Psi_A\big\rangle\Big\vert^2 \, ,
\label{FSIdist}
\end{equation}
where the operator $S(1,...A)$  describes interactions between the
struck proton
and
the remaining (A-1) spectator nucleons.
In the diffractive regime, the Glauber model gives
\begin{equation}
S(1,...A)\ =\ \prod_{n=2}^A\Bigg\{1 - \Gamma(\vec{b}-\vec{b}_{n})
\theta(z_n - z)
\Bigg\}\, .
\end{equation}
Here we decompose $\vec{r}_{i}=(\vec{b}_{i},z_{i})$, taking the $z$
axis along $\vec{q}$.  The profile function of the $pN$ interaction,
$\Gamma(\vec{b})$, is usually parameterized as
\begin{equation}
\Gamma(\vec{b})\ =\ {{\sigma_{tot}(1-i\rho)}\over{4\pi b_o^2}}
\exp\Big(-{{\vec{b}^2}\over{2 b_o^2}}\Big)\, .
\label{gam}
\end{equation}
Here $\rho$ is the $Re/Im$ ratio for the forward elastic $pN$
scattering amplitude, $b_{o}^{2}$ is the diffraction slope.  In the
GeV energy range, $\sigma_{tot}$ $\approx$ 40 mb, $\rho$ $\approx$
0.3$\div$0.4, $b_o$ $\approx$ 0.5 fm \cite{lasi,lll93}.  For the
deuteron target, the FSI distortion factor takes on the particularly
simple form
\begin{equation}
S(\vec{r})\ =\ 1\ -\ \Gamma(\vec{b}) \theta(-z).
\label{sop}
\end{equation}
Here $\vec{r}$ is the proton-neutron separation. Eq.~(\ref{sop})
illustrates basic features of FSI at high energy and momentum transfer
as reflected in the Glauber formalism: (i) The $\theta(-z)$ tells that
FSI is possible only provided that the spectator nucleon was in the
forward hemisphere with respect to the struck proton.  (ii) The form
of $\Gamma(\vec{b})$ tells that the struck proton wave is distorted
only at small $|\vec{b}| \lsim b_{o}$.  (iii) Because $\rho$ is small,
$\Gamma(\vec{b})$ is dominated by the imaginary part of the $p$-$n$
elastic scattering amplitude and the distortion factor $S(\vec{r})$
predominantly gives an absorption of the struck proton wave at small
$\vec{b}$. At lower energies, the term "inelastic event" indicates
that the target nucleus breaks up, whereas at the considered high
four-momentum transfer $(e,e'p)$ experiments, "inelastic" means that
the proton breaks up when undergoing FSI with the residual nucleus.
The fact that $\rho$ is small and $\sigma_{el} < \sigma_{in}$ means
that those inelastic events are dominating at GeV energies.  In
contrast to FSI at low energies, which is S-wave dominated and
therefore isotropic, the above basic properties of FSI at high
energies demonstrate that the transverse and longitudinal directions,
and the forward and backward hemisphere, have different roles, with
the phenomenological consequences of a marked angular anisotropy, and
forward-backward asymmetry.

In the Glauber model the FSI factor has no free parameters, it is
fully specified in terms of the free nucleon scattering amplitude. At
the energies which are relevant for this work, the Glauber model
description of hadron-nucleus scattering is well tested
\cite{abv78,Wallace}.

The above expression has been used for calculating the longitudinal
response in (e,e'p) on deuteron \cite{Deuter}, with realistic Bonn
\cite{bonn} and Paris \cite{paris} wave functions. For the $^{4}He$
target, one must use a wave function with realistic Jastrow-type
correlations:
\begin{equation}
\Psi\ =\ \prod_{i<j}(1-F_{ij})\Psi_o.
\label{wfun}
\end{equation}
Here $\Psi_o$ = $\exp\{-(r_1^2+r_2^2+r_3^2+r_4^2)/2R_o^2\}$ is a
harmonic oscillator mean field wave function, and $1-F_{ij}$ = $1-C_o
\exp(-r_{ij}^2/2r_c^2)$ is a correlation operator expressing hard
($C_o$=1) or soft ($C_o<1$) core repulsion when two nucleon centers
are within a distance $r_c$ (the practical calculations are done in
Jacobi coordinates).  This function contains the dominating features
of the $^4$He ground state, with two exceptions, namely 3-body and
d-wave correlations (see e.g. \cite{wiringa92}). However, we show that
the sensitivity towards those corrections tends to be lost when FSI
are included.  In the Jastrow-type correlation $F_{ij}$, we take a
standard value of $r_{c}=0.5$ fm. Then, the choice $R_o$ = 1.29 fm
correctly reproduces the experimental charge radius (taking into
account the finite nucleon size \cite{devries}).  The qualitative
agreement of our PWIA distribution with the results of the Monte Carlo
calculation of \cite{wiringa92} and the parametrization given by
\cite{ciofi95} is satisfactory.  An extensive discussion of the
results of the full calculation of the distribution (\ref{FSIdist})
with the wave function (\ref{wfun}) is presented elsewhere
\cite{NewHe4}. Here we only wish to compare certain common features of
the deuteron and $^4$He (e,e'p) distributions at large $p_{m}$.

A new insight into FSI effects is needed in the regime of diffractive
$N$-$N$ scattering.  At lower energies, FSI act like a correlated
response of the full residual nucleus to the passage of the
ejectile. The characteristic parameter of FSI is the nuclear radius.
In the GeV energy range, the wavelength of the struck proton is short
and the crucial parameter is the radius of diffractive $pN$
scattering, which is much smaller than the nuclear radii, $b_{0}\ll
R_{A}$, and approximately equal to the correlation radius,
$b_{0}\approx r_c$. For this reason one can expect a large FSI
contribution at large $p_m$ $\sim$ $1/b_{0}$ $\approx$ $1/r_c$. For
instance, for transverse $p_m$, the FSI factor (\ref{gam}) gives rise
to a large Fourier transform in (\ref{FSIdist}) and to a large
$p_{m\perp}$ tail of the missing momentum distribution
$W(\vec{p}_{m})$:
\arr W(\vec{p}_{m}) \propto \left|\int d^{2}\vec{b}
\Gamma(\vec{b}) \exp(i\vec{p}_{\perp}\vec{b})\right|^{2} = 4\pi
{d\sigma_{el} \over dp_{\perp}^{2}} = {1 \over
4}\sigma_{tot}^{2}(1+\rho^{2})\exp(-b_{o}^{2}p_{\perp}^{2}) \, .
\label{Pperp}
\endarr
This contribution to $W(\vec{p}_{m})$ can be attributed to
elastic rescattering of the struck proton on spectator nucleons.  To a
crude approximation, the $p_{m\perp}$ dependence in (\ref{Pperp}) does
not depend on the target nucleus, as $b_o^2 << R_A^2$. In the absence
of FSI, both in deuteron and in larger nuclei the large $p_m$ tail of
the single particle distribution would have been dominated by
short-distance nucleon-nucleon interactions in the nuclear ground
state.  However our results \cite{Helium4,Deuter} show that this is
not the case when FSI are included.

In Fig.~\ref{fig1}a we show the PWIA and the full transverse momentum
distribution $W(\vec p_m)$ including FSI for the deuteron wave
functions calculated from the Bonn and Paris models. The well known
smaller D-wave content of the Bonn model makes the corresponding PWIA
much smaller than in the Paris model at large $p_m$.  This difference
gives an estimate of the uncertainties in the predictions of modern
theories of the $NN$ interaction at large $p_m \gsim 1.5 fm^{-1}$.
However, when FSI are included the differences between the two
distributions disappear: FSI, which distort the struck proton's wave
function only at small distances $|\vec b| \lsim b_o$, are hardly
sensitive to the deuteron D-wave (where the centrifugal barrier keeps
nucleons apart).  In Fig.~\ref{fig1}b we show how the distribution
$W(\vec p_m)$ for the $^4He(e,e'p)$ reaction is changed for different
types of correlations, going from $C_o=0$ (pure mean field) to $C_o =
1$ (hard core).  For transverse kinematics, the FSI-dominated
distributions are not very sensitive to such drastic changes in the
ground state wave function. From Fig.~\ref{fig1} we learn that in the
leading order the FSI are not too sensitive to the large-$k$
components of the nuclear ground state $\Psi_A$.  In transverse
kinematics, the FSI redistribute strength from small and intermediate
momenta to the large $p_m$ tail of $W(\vec p_m)$, and therefore the
transverse missing momentum distribution including FSI effects is
mainly sensitive to the bulk of the nuclear ground state (in deuteron
the S-wave, in complex nuclei the mean-field orbitals).

In Fig.~\ref{figmdcomp}, we come to the main point: in transverse and
parallel kinematics (both forward and backward) we show the
distributions (including FSI) for $^4$He and for the deuteron. The
latter is multiplied by 3 to take into account that in $^4$He we have
3 FSI scatterers.  The similarity between the FSI-inclusive
distributions for $^4$He and deuteron at $p_m$ larger than 1.5
fm$^{-1}$ is impressing, especially in transverse kinematics.  It is
even more striking if one compares the transverse and longitudinal
distributions corresponding to one and the same nucleus (either
deuteron or $^4He$). They are equal in the fully isotropic PWIA
prediction.  On the contrary, at large $p_m$ the full distributions
$W(\vec p_m)$ differ by one order of magnitude, attesting the FSI
dominance.  So, at last, universality appears, but it is driven by FSI
rather than by an intrinsic feature of the nuclear ground state wave
function.

In transverse kinematics, the PWIA curve is overwhelmed by the FSI
effect by orders of magnitude. There is no hope to find $direct$
quasi-deuteron effects in experiments performed in $transverse$
kinematics. Even more confusion can arise from experiments in which
events which are taken at different angles $\theta$ of the missing
momentum $\vec p_m$ are put together as a function of $|\vec p_m|$ and
are not presented in separate distributions.  In transverse kinematics
the FSI dominance is so marked that it seems difficult to think that a
more refined $^4He$ ground state can modify this situation, especially
as the transverse missing momentum distribution for the $D(e,e'p)$
reaction calculated with realistic wave functions is also dominated by
FSI and as the FSI effects in $^4He$ should be stronger than in the
rather dilute deuteron.

In longitudinal kinematics, the situation seems more interesting.
There, our results suggest that PWIA and FSI effects are in
competition, and a quantitative understanding of their interplay can
become decisive.  At these kinematics, at large $p_{m,z}$, FSI effects
are mainly due to the $\theta(-z)$ factor in (\ref{sop}). The related
discontinuity introduces high-momentum longitudinal components.  This
is again a universal property of the FSI, the presence of which does
not depend on the specific target nucleus, although there is a slight
sensitivity to the form of the correlation function.

Apart from this large $p_{m,z}$ tail, the interference between the
PWIA amplitude and the $\propto \rho$ component of the FSI amplitude,
leads to a forward-backward (F/B onwards) asymmetry, shown in
Figs.~\ref{figfba}a,b for $^4He$ and $D$.  The mere presence of a
large F/B asymmetry suggests strong FSI effects, and again its $p_m$
dependence for deuteron and $^4He$ with soft core/hard core
correlations included is qualitatively similar. The realistic deuteron
wave functions do already include effects of the short range $NN$
interaction.  However, as the F/B asymmetry is a PWIA-FSI interference
effect, it also contains some information on the nuclear ground state,
as was shown in detail in \cite{tensor} for the special case of
polarized deuteron, and therefore and due to the different values of
$\rho$ for $D$ and $^4He$ the similarity is less pronounced than in
transverse kinematics.

Concluding, we have shown that FSI create strong similarity patterns
in the energy-integrated momentum distributions for (e,e'p) on $^4He$
and deuteron at large missing momenta.  This will allow for detailed
studies of universal features of nuclear FSI in the diffractive
regime, but even make testing short range structures in the nuclear
ground state much more difficult.  In particular, in transverse
kinematics this task seems hopeless. It has to be stressed that FSI
effects cannot be described by a simple overall renormalization factor
as they depend strongly on the specific kinematics.  A further
analysis of the situation in longitudinal kinematics requires more
sophisticated models for the $^4He$, due to the complex interplay of
FSI and PWIA effects there.

{\bf Acknowledgments:} This work was done during a series of visits of
A.B. at IKP, KFA J\"ulich (Germany), and of S.J. at Pavia University
(Italy), supported by IKP and by the ``Vigoni'' program of DAAD
(Germany) and the Conferenza Permanente dei Rettori (Italy).
A.B. thanks J.Speth for the hospitality at IKP and S.J. thanks S.Boffi
for the hospitality at the University of Pavia. This work was also
supported by the INTAS Grant No. 93-239.


\newpage
{\large \bf Figure Captions}
\vspace{0.5cm}
\begin{figure}[h]
\caption{a) The PWIA missing momentum distribution $N(p_m)$ in the
$D(e,e'p)$ reaction calculated with the Bonn wave function \protect
\cite{bonn} (dashed line) and with the Paris wave function \protect
\cite{paris} (dotted line) and the full missing momentum distribution
$W(\vec p_m)$ including FSI in the $D(e,e'p)$ reaction at $\theta =
90^o$ calculated with the Bonn wave function (solid line) and the
Paris wave function (dash-dotted line).  b) The full missing momentum
distribution $W(\vec p_m)$ including FSI in the $^4He(e,e'p)$ reaction
for hard core correlations $C_o = 1$ (solid line), soft core
correlations $C_o = 0.5$ (dashed line), and pure mean field, $C_o=0$
(dotted line).  }
\label{fig1}
\end{figure}
\vspace{-0.6cm}
\begin{figure}[h]
\caption{The full missing momentum distribution $W(\vec p_m)$ for the
$^4He(e,e'p)$ reaction (solid line) and for the $D(e,e'p)$ reaction
(dashed line) multiplied with a factor of $3$ to account for the
higher number of FSI scatterers in $^4He$. For comparison, the PWIA
missing momentum distribution $N(p_m)$ for $^4He$ is also shown
(dotted line). On top, we show transverse kinematics, $\theta = 90^o$,
and in the middle and the lower panel we show parallel and
antiparallel kinematics, $\theta = 0^o$ and $\theta = 180^o$.}
\label{figmdcomp}
\end{figure}
\vspace{-0.5cm}
\begin{figure}[t,h]
\caption{The forward-backward asymmetry $A_{FB} = \frac
{W(\theta=0^{o};p_{m})-W(\theta=180^{o};p_{m})}
{W(\theta=0^{o};p_{m})+W(\theta=180^{o};p_{m})}$ is shown for the
reaction $D(e,e'p)$ in a) and for the reaction $^4He(e,e'p)$ in
b). For $^4He$, different correlations were used: hard core, $C_o =
1$, (solid line), soft core, $C_o = 0.5$, (dashed line), and pure mean
field, $C_o = 0$ (dotted line).}
\label{figfba}
\end{figure}

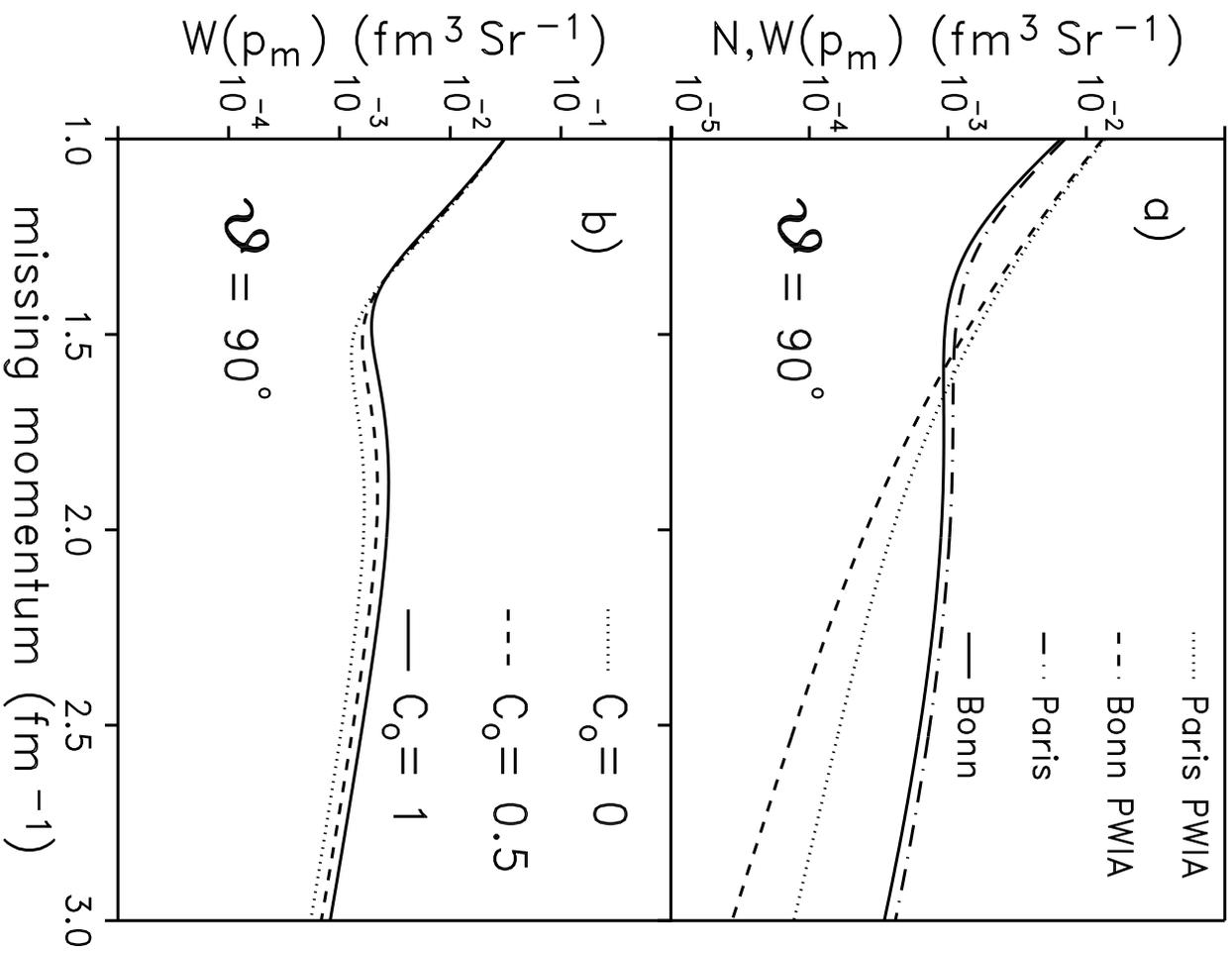
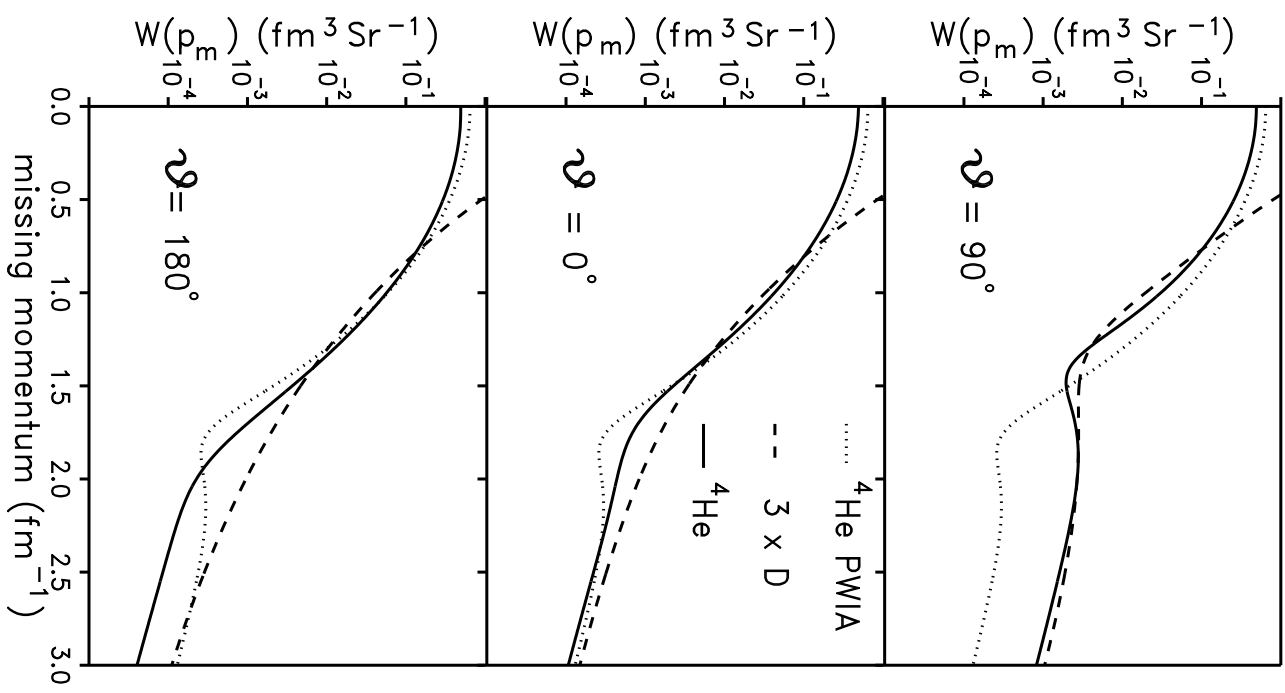
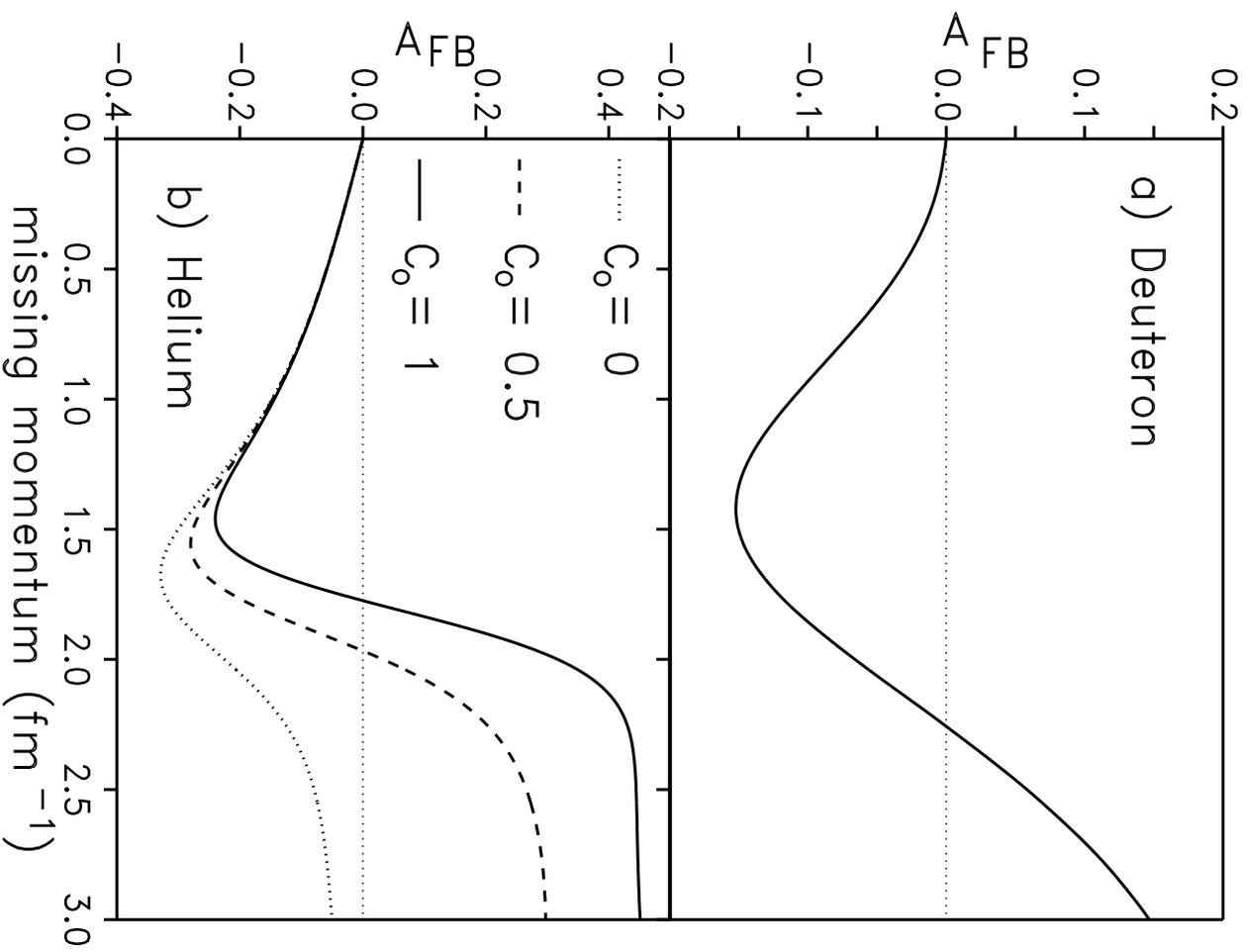
\end{document}